\documentclass[a4paper,11pt]{article}
\usepackage[T1]{fontenc}
\usepackage[utf8]{inputenc}
\usepackage{lmodern}
\usepackage{graphicx}
\usepackage{epstopdf} 
\usepackage{color}
\usepackage{bbold}
\usepackage{amsmath}
\title{A BRST Characterization of the Broken Phase Observables of the Confining Complex Theory}

\author{R. L. P. G. Amaral$^{a}$\footnote{email: rubens@if.uff.br} ,
	V. E. R. Lemes$^{b}$\footnote{email: verlemes@gmail.com},
	O. S. Ventura$^{c}$\footnote{email: ozemar.ventura@cefet-rj.br}, L.C.Q.Vilar $^{b}$\footnote{email: lcqvilar@gmail.com}  \\
	\small \em $^a$Instituto de F\'{\i}sica, Universidade Federal do Fluminense\\
	\small \em Av. Litor\^anea S/N, Boa Viagem, Niter\'oi-RJ CEP. 24210-340,
	Brazil\\
	\small \em $^b$Instituto de F\'\i sica, Universidade do Estado do Rio de
	Janeiro,\\
	\small \em Rua S\~{a}o Francisco Xavier 524, Maracan\~{a}, Rio de Janeiro - RJ,
	20550-013, Brazil\\
	\small \em $^c$Departamento de F\'\i sica, Centro Federal de Educa\c{c}\~ao Tecnol\'ogica do Rio de
	Janeiro\\
	\small\em Av.Maracan\~a 249, 20271-110, Rio de Janeiro - RJ, Brazil}

\begin{document}
	\maketitle
	\begin{abstract}	
	Some time ago we have introduced a route to provide confinement in the sense that particle excitations would appear from condensates of fields that do not have physical asymptotic states \cite{ALVV20}. We envisaged this mechanism in an asymmetric vacuum phase of a complex gauge field theory. More recently, we showed how to define a BRST operator to the broken phase of a generic spontaneous broken field theory \cite{ALVV22}. Our intention here is to apply these late concepts to the previous complex field theory in order to give a cohomological characterization of those condensed states as actual physical observables of this theory in its broken phase.

\end{abstract}
	
\section{Introduction}

In a recent work \cite{ALVV22}, we addressed the theme of the BRST characterization of observables in the broken phase of a spontaneous symmetry breaking of a grand unified GUT theory. This was a question which imposed itself in an obvious way, as new observables (as gauge field masses or independent couplings) are the main objective of any spontaneous symmetry breaking theory, and the BRST description of these observables was still lacking. We accomplished this goal by initially  understanding that the symmetry breaking process demands a redefinition of the BRST symmetries taking into account the displacement of the vacuum in the broken phase. The first outcome of this reasoning leads to a nilpotent operator (called $s_q$ in \cite{ALVV22}) adapted to the new vacuum. This operator appears when we isolate the scalar degrees of freedom that are encompassed by the gauge fields acquiring masses in the breaking process. This system of symmetries is already sufficient to characterize the new observables associated to new non-trivial invariant objects of the $s_q$ cohomology. However, $s_q$ alone is not able to link the renormalization of the coupling constants in order to reach a unique value at the symmetry breaking scale, as is demanded by the fact that we intend to describe a GUT phase transition based on a simple group. This means that we must add new elements in such a way that as long as the energy increases close to the breaking scale the running couplings may converge to a common value. These elements were called $\delta$ in \cite{ALVV22}. The complete symmetry operator $s_q + \delta$ is again nilpotent. It becomes essential to the definition of the 't Hooft gauge in this regime, and in this sense we now interpret $s_q + \delta$ as the BRST operator of the broken phase. Then, the role played by $\delta$ seems to be just a constraint on the dynamics of the broken theory in order to recover the symmetric theory at the breaking scale. It still preserves the non-triviality of the main cocycles already described by the cohomology of $s_q$, i.e., the independence of the couplings and gauge field masses are not disturbed by $\delta$ \cite{ALVV22}.

This issue becomes relevant as we intend to apply this development in the context of a phase transition process now leading to confined states. Once an analogous $s_q$ operator is defined, we envisage its use in the search of possible condensate states which cannot be defined in any other way. With this intent, let us describe the field theory with a spontaneous symmetry breaking with attainable confined states in the broken phase.

This theory was presented in our work \cite{ALVV20}. There we argued that the previous progress of Gribov`s ideas on confinement \cite{Gribov} (in the same context, the Gribov-Zwanziger/GZ theory
\cite{Zw12,Zw89,Dud08}) was naturally leading the research towards theories where the loss of positivity should take a prominent role. Regarding this point, we highlight first the definition of the concept of i-particles, fundamental elements from which condensates satisfying the propagator positivity criteria \cite{100,101,102,103} would be possible \cite{Bau10, Cap11, Dud10, Dud11}. Then, when the construction of such condensates failed in the standard GZ scenario, a new idea emerged 
associating fundamental fields directly to i-particles, the so called replica model \cite{Cap11,Sor10}. In this work, the loss of positivity appears as an inevitable outcome of Gribov´s approach to  gluons confinement. More than this, the replica model contains a complex gauge field hidden in its structure, and this was the guideline to our research. Another
point to be considered here is that, although the GZ model can be obtained from the spontaneous symmetry breaking of a trivial BRST term \cite{Vil11,Sc15}, following the former proposal contained in \cite{Fj83}, this is not enough to build i-particles. The missing ingredient is the breakdown of positivity, naturally embedded within complex gauge theories. We joined these ideas and studied a complex gauge field theory symmetric under a complex group that undergoes a spontaneous symmetry breaking process \cite{ALVV20}. In the asymmetric phase, the propagators of the gauge fields associated to the broken directions become i-particles. Consequently, these fundamental fields are not associated  to asymptotic particle states anymore, but a positive norm subspace can still be reached if we look at condensates formed from these fields. Such objects should propagate respecting K\"all\'{e}n-Lehmann (KL) spectral representation, and then would define candidates for observable states \cite{Dud11}. In \cite{ALVV20}, we succeeded in presenting a possible observable meeting these requisites.

On the other hand, if we expect this object to be a physical observable, it should be characterized in the BRST cohomology. This means that it must be invariant under the action of the nilpotent BRST operator and at the same time it should not be written as a BRST variation (see \cite{Sor95} for an introduction on the BRST renormalization). At the time when \cite{ALVV20} was written, we did not have the BRST scheme developed to the broken phase. Then we based our study on the non-trivial objects of the symmetric phase, where the BRST operator was well known, and hope that among them we could find at least an indication of a condensate. In fact, this is one of the results shown in \cite{ALVV20}. Now, with the realization of the BRST symmetry for the asymmetric vacuum described in \cite{ALVV22}, the $s_q$ operator mentioned above, we can return to the complex field theory and study in detail its broken phase. We will show that our hint on the condensate of \cite{ALVV20} is justified, as the part of the invariant object of the symmetric phase that has a KL propagation is actually a non-trivial cocycle of this new broken phase BRST operator. Also, with this machinery, we will be able to identify new contributions for this broken phase observable. Finally, it is important to highlight a point discussed in \cite{ALVV20}
that our model allows the 
confinement of quarks and gluons in the same theory. For the quarks, we considered the
confining potential criterion of 't Hooft
\cite{tH03, tH003, tH07}, which led to the the
linear Cornell potential \cite{400,401,402}.

In the sequence, in Section 2 we review the complex gauge theory, and the conceptual developments
that lead to it, and in Section 3 we construct the BRST
operator $s_q$ for its broken phase. In Section 4 we show how this reveals the loss of
holomorphicity in this phase, a fact that we argue to be behind the emergence of the gluon
condensates which we present and characterize as observables in the cohomological sense.
Finally, our conclusions are shown in Section 5.

\section{The Complex Gauge Theory}

Before reviewing our complex gauge theory, let us give some context of the ideas that led us to the development of \cite{ALVV20}. 

The original implementation of the GZ theory had the basic structure of a trivial BRST element added to a Yang-Mills lagrangian, with the intent that gluon confinement would ultimately be perceived as an inevitable outcome of pure Yang-Mills theory.  However, in order to create dynamics from this topological structure, a process of fixation of BRST sources is embedded in Zwanziger's scheme \cite{Zw93}. This leads to a soft breaking of the BRST invariance \cite{So09}. In fact, the instability of the BRST symmetry in the non-perturbative regime was understood a long time ago \cite{Fj83}.  Although the renormalizability of the GZ theory has been checked in the Landau gauge \cite{Ted11}, it soon became clear that this process implied a gauge dependent construction \cite{Lav11, Lav12}. Each gauge fixing would require a different horizon function, which would intrinsically carry this gauge dependence \cite{Lav15}. 

The acknowledgement of this inconsistency led to the recovery of an older interpretation of the full GZ theory as a spontaneous symmetry breaking of BRST starting from pure Yang-Mills plus the trivial topological sector \cite{Mag94}. The mechanism itself generating this phase transition is not described, but the broken phase is characterized by the existence of a non null vacuum expectation value of a BRST trivial element. Some problems with the derivation of the GZ theory from this symmetry breaking point of view were then pointed out, mainly from the fact that the whole approach depended explicitly on the space time coordinate \cite{Dud08}. Since then, this criticism was circumvented by different approaches.  The implementation of a finite volume quantization was carried out in \cite{Zw12, Sc15}, in a way to avoid the inconsistency of the original Maggiore-Schaden construction. And another spontaneous symmetry breaking effect, also free from the explicit space time dependence, again by the extension of the field content of the theory, was presented in \cite{Sor12}. Nevertheless, these options were only developed for the specific Landau gauge (as it was actually recognized in \cite{Sor12}), and so may suffer from the same gauge dependence problem that motivated the analysis in \cite{Lav11, Lav12} (also the existence of preferred directions on the resulting vacuum seems to be another drawback).

These developments, although not successful, brought us the impression that GZ could in fact be describing a phase of a larger theory, when it is already confined. At the same time, it is natural to expect that a process of spontaneous symmetry breaking should play a major role, since this is the mechanism that we find in theoretical physics that can make this transition and at the same time preserve renormalizability and unitarity. 

On the other hand, as it is cleared stressed in \cite{Sor12}, the preservation of a BRST symmetry does not guarantee that a unitary description will be reached. One should in the end verify if the theory allows for a positive norm subspace of the confined degrees of freedom. As in a confined phase, the elementary excitations are expected not to have asymptotic states, this subspace must be searched among the two point correlators of composite operators constructed from the basic fields. In fact, it is very hard to describe correlators with this property in the GZ theory. Several developments lead to the foundation of the important concept of i-particles, fundamental elements from which condensates satisfying positivity criteria would be possible \cite{Bau10, Cap11, Dud10, Dud11}. Propagators of i-particles would be behind the formation of Gribov propagators. However, the fact that the fundamental fields in GZ do not precisely represent i-particles led to the impossibility of defining condensates with the necessary properties to describe physical observables in the theory \cite{Zw89}. A new idea to overcome this obstacle was the further development of the replica model  \cite{Sor10}. Created in order to associate fundamental fields directly to i-particles, it showed once more the loss of positivity as an inevitable outcome in this confinement scenario. But another ingredient appeared: a complex gauge field is hidden in the replica model. 

Then, we take complex gauge field theory as an appropriate environment to the description of i-particles. As already mentioned, we will be led to the conclusion that fundamental fields are not associated to asymptotic particle states. This can be seen as a precept to the confinement, but the main issue is that we must recover the physical spectrum of excitations of the theory, or else such theory will remain physically meaningless. A possible path opens if we assure the possibility of defining condensates from such fields. In fact, the concept of i-particles is born inside the GZ theory as a building block for condensates. These would be formed from vertices joining simultaneously pairs of i-particles and anti-i-particles. Objects built in this way would propagate respecting K\"all\'{e}n-Lehmann spectral representation, and then would define candidates for observable states. And according to what is derived in the replica model, this combination in pairs of i-particles and anti-i-particles is essential for the success of the construction.

Our theory begins with the usual transformation of a complex gauge field $ \mathcal{A}_{\mu }$ in the adjoint representation

\begin{equation}
\mathcal{A}_{\mu } \longrightarrow  \mathcal{A}_{\mu }^{'} = G^{-1} \mathcal{A}_{\mu } G + {i \over g} G^{-1}(\partial_{\mu } G) \label{agaugetransf}
\end{equation}
and, in the case of a complex group, we have the possibility of defining a conjugated field  $\bar{\mathcal{A}}_{\mu }$ transforming  distinctly as

\begin{equation}
\bar{\mathcal{A}}_{\mu } \longrightarrow  \bar{\mathcal{A}}_{\mu }^{'} = G^{\dagger} \bar{\mathcal{A}}_{\mu } G^{\dagger -1} + {i \over g} G^{\dagger} (\partial_{\mu } G^{\dagger -1}) 
\label{abargaugetransf}
\end{equation}
since in a complex group $G^{-1} \neq G^{\dagger}$. In the next step , we can define covariant curvatures $\mathcal{F}_{\mu \nu}$ and $\bar{\mathcal{F}}_{\mu \nu}$ for ${\mathcal{A}}_{\mu }$ and $\bar{\mathcal{A}}_{\mu }$ respectively. Thus we immediately  see that invariant objects built exclusively from these curvatures will be holomorphic in the sense that once ${\mathcal{A}}_{\mu }$ is found in a monomial, it will not contain $\bar{\mathcal{A}}_{\mu }$, and vice-versa. For example, we will be able to define curvature invariants as $ Tr \mathcal{F}^2$ and $ Tr {\bar{\mathcal{F}}}^2$. Therefore the inevitable loss of positivity. Positivity would only be ensured in the non-holomorphic element $ Tr \mathcal{F} \bar {\mathcal{F}} $, which is not invariant by the complex gauge transformations (\ref{agaugetransf}) e (\ref{abargaugetransf}). Naturally, an i-particle will be associated to the ${\mathcal{A}}_{\mu }$ field, and an anti-i-particle to $\bar{\mathcal{A}}_{\mu }$. 

We also assumed that the GZ theory should be embedded in a larger theory with a spontaneous symmetry breaking sector. The usual option is to add a scalar sector to the theory. The first idea would be to work with a complex scalar field and its conjugate, which would take us to a holomorphic scalar sector. This route leads to difficulties in the definition of gluon and fermion condensates. But there is an alternative path.

This comes from the freedom that the complex group gives us. The fact that $G^{-1} \neq G^{\dagger}$ allows us an unorthodox  proposition for an adjoint inspired  scalar field transformation

\begin{equation}
\phi \longrightarrow  \phi^{'} = G^{\dagger} \phi G \label{phigaugetransf}
\end{equation}
and also
\begin{equation}
\psi \longrightarrow  \psi^{'} = G^{-1} \psi G^{\dagger -1} \, .\label{phibargaugetransf}
\end{equation}

These scalar fields, in isolation, do not form invariants. But joining them together we find $Tr \phi \psi $, which is invariant under this action of the complex group. 

Here we need to take some care to understand what is implied in the transformations (\ref{phigaugetransf}) and (\ref{phibargaugetransf}). In these it is implicit a left and a right action on the field, which indicates that the field carries a representation of the algebra, as it occurs in the traditional adjoint representation. However, as the transformations in (\ref{phigaugetransf},\ref{phibargaugetransf}) involve $ G^{\dagger}$ and not $G^{-1} $, we can conclude that for a general group these transformations are not closed inside the vectorial space of the algebra. This means that the admissibility of this pattern of transformation will depend on specific choices of the gauge group. Since  $G^{-1} \neq G^{\dagger}$, we know that unitary groups are not eligible. Nevertheless, the transformations (\ref{phigaugetransf},\ref{phibargaugetransf}) preserve hermiticity, $i.e.$, if $\phi $ and  $ \psi $ are hermitians, then so will be $\phi ' $ and  $ \psi ' $. An hermitian basis for the algebra is characteristic of the real unitary groups, but in the complex extension such basis is acceptable for the $SL(N,C)$ groups. For instance, in the case of the complex $SL(2,C)$, we can take the Pauli matrices as the generator basis, or Gell-Mann matrices for the complex $SL(3,C)$. So, let us assume that we are working with a complex $SL(N,C)$ as the gauge group. Even so, it is not still warranted that in (\ref{phigaugetransf},\ref{phibargaugetransf}) a hermitian matrix of the $sl(N,C)$ algebra will be rotated into another algebra matrix by the action of the $SL(N,C)$. The minimal cost to obtain a consistent construction is to impose that such matrices of the  $sl(N,C)$ algebra used to define the scalar fields belong to the fundamental representation. In this case, we will have a basis with  $N^2 -1$ $N \times N$ matrices, and consistency will be achieved in (\ref{phigaugetransf}) and (\ref{phibargaugetransf}) if we suppose the existence of a further $\phi^0 $ component associated with the $N \times N$ identity (and the same for $ \psi $).

We used these building blocks in the construction of the complex gauge theory presented in \cite{ALVV20}. The action has the form
\begin{eqnarray}
S= \int d^{4}x (\frac{i}{4}\mathcal{F}^{a}_{\mu\nu}\mathcal{F}^{a}_{\mu\nu}-\frac{i}{4}\bar{\mathcal{F}}^{a}_{\mu\nu}\bar{\mathcal{F}}^{a}_{\mu\nu} + Tr(\mathcal{D}_{\mu}\varphi)(\mathcal{D}_{\mu}\psi)  + V(\varphi,\psi ) + S_{GF} ) \, .
\label{action}
\end{eqnarray}

In this action we find the complex gauge field $\mathcal{A}_{\mu }$ and its associated ghost $c$ transforming in the usual way under BRST
\begin{eqnarray}
s\mathcal {A}_{\mu }&=&-(\partial_{\mu }c-ig[\mathcal{A}_{\mu },c]) ,\nonumber \\
sc &=& -igc^{2}; \label{sA}
\end{eqnarray}
and the complex conjugated gauge field $ \bar{\mathcal{A}}_{\mu}$ and its associated ghost $\bar{c}$ transforming as 

\begin{eqnarray}
s\bar{\mathcal {A}}_{\mu}&=&-(\partial_{\mu }\bar{c}-ig[\bar{\mathcal{A}}_{\mu},\bar{c}]) , \nonumber \\
s\bar{c} &=& -ig\bar{c}^{2} . \label{sAbar}
\end{eqnarray}
 With them we define the usual curvature $\mathcal{F}_{\mu \nu}$ associated to $\mathcal{A}_{\mu }$
\begin{eqnarray}
\mathcal{F}_{\mu \nu}(\mathcal{A})=\partial_\mu \mathcal{A}_\nu - \partial_{\nu } \mathcal{A}_\mu -ig[\mathcal{A}_\mu, \mathcal{A}_\nu],
\label{F}
\end{eqnarray}
and its complex conjugate $\bar{\mathcal{F}}_{\mu \nu}$
\begin{eqnarray}
\bar {\mathcal{F}}_{\mu \nu}(\bar{\mathcal{A}})=\partial_\mu \bar{ \mathcal{A}}_\nu - \partial_{\nu } \bar {\mathcal{A}}_\mu -ig[\bar {\mathcal{A}}_\mu, \bar {\mathcal{A}}_\nu].
\label{Fbar}
\end{eqnarray}
The set of transformations (\ref{sA}, \ref{sAbar}) implies that only holomorphic elements as $ Tr \mathcal{F}^2$ and $ Tr {\bar {\mathcal{F}}}^2$ are invariant. Once more, the non-holomorphic element $ Tr \mathcal{F} \bar {\mathcal{F}} $ is not allowed by these BRST transformations.

 A distinct feature of the model is the pair of scalar fields $\varphi$ and $\psi$ transforming as
\begin{eqnarray}
s\varphi &=& ig\varphi c -ig\bar{c}\varphi  , \nonumber \\
s\psi &=& ig\psi \bar{c}-ig c\psi  , \label{spsi}
\end{eqnarray}
following (\ref{phigaugetransf}, \ref{phibargaugetransf}) and allowed by the complex group structure. In \cite{ALVV20} we chose  $SL(3,C)$ as the complex gauge group, so that the scalar field $\varphi=\varphi^{A}T^{A}$ (and analogously $\psi$)  has components from $A=0$ to $A=8$, with $T^{A}$,  $A=1,...,8$, representing the eight Gell-Mann matrices, and $T^{0}=\sqrt{ \frac{1}{6}} I$, with $I$ the 3x3 identity matrix. The gauge fields ${\mathcal{A}}_\mu$ and $\bar{{\mathcal{A}}}_\mu$ then have eight complex components each, projected also on the eight $T^{A}$ Gell-Mann matrices.

 From them we define the covariant derivatives

\begin{eqnarray}
\mathcal{D}_{\mu } \varphi &=& \partial_{\mu }\varphi + ig\varphi \mathcal{A}_{\mu }-ig \bar{\mathcal{A}}_{\mu }\varphi ,\nonumber \\
{\mathcal{D}_{\mu } }\psi &=& {\partial_{\mu }\psi}+ig \psi \bar{\mathcal{A}}_{\mu }-ig \mathcal{A}_{\mu }\psi , \label{nablapsi}
\end{eqnarray}
the sense of covariance being understood by their transformations
\begin{eqnarray}
s\mathcal{D}_{\mu } \varphi &=& ig(\mathcal{D}_{\mu } \varphi ) c -ig \bar{c}(\mathcal{D}_{\mu } \varphi ) , \nonumber \\
s{\mathcal{D}_{\mu } }\psi&=&ig({\mathcal{D}a_{\mu } \psi})\bar{c}-ig c({\mathcal{D}_{\mu } \varphi}). \label{snablapsi}
\end{eqnarray}
Such special derivatives in (\ref{action}) are responsible for the i-particle generation after the symmetry breaking, together with the production of an inter-quark confining potential in the broken phase if we couple this system to fermions \cite{ALVV20}.

Regarding the gauge fixing $S_{GF}$, we remember that in a symmetry breaking process, the adequate gauge fixing is the 't Hooft gauge. Then, to implement this we introduced two gauge conditions
\begin{eqnarray}
G&=& \partial_{\mu}\mathcal{A}_{\mu} +\frac{g\alpha}2\left(\psi\mu-\mu\varphi-\frac13Tr\left\{\psi\mu-\mu\varphi\right\} I \right)\nonumber \\
\bar G&=& \partial_{\mu}\bar{\mathcal{A}}_{\mu}+\frac{g\alpha}2 \left(\mu\psi-\varphi \mu-\frac 13Tr\left\{\mu\psi-\varphi \mu\right\} I \right),
\label{Gcondition}
\end{eqnarray}
where $\alpha$ is a gauge parameter,  and a pair of anti-ghosts $q$ and $\bar{q}$, and their respective Lagrange multipliers $b$ and $\bar b$, transforming in BRST doublets

\begin{eqnarray}
 s q= -ib, &&  s b=0, \nonumber \\
 s\bar{q}= i\bar b, &&  s \bar b=0.
 \label{antighosts}
\end{eqnarray}
Then, our gauge fixing took the form

\begin{equation}
S_{GF}= s\int d^{4}x \left( Tr\left(-2q{G} -2\bar q\bar G+\alpha  qb +\alpha \bar{q}\bar b\right)\right).
\label{GF}
\end{equation}

Finally, $V(\varphi, \psi )$ was designed in analogy with a standard scalar quartic potential

\begin{eqnarray}
V(\varphi, \psi )= -\frac{m^{2}}{2} \varphi^{A} \psi^{A}+\frac{\lambda }{4}(\varphi^{A} \psi^{A})^{2}   ,
\label{potential}
\end{eqnarray}
which attains a local minimum at
\begin{equation}
<\varphi^{A} \psi^{A}>= \frac{m^{2}}{\lambda }.
\label{minima}
\end{equation}

In \cite{ALVV20} we chose the following vacuum expectation values vev for these scalars
\begin{eqnarray}
\varphi&\mapsto& \varphi + \mu \nonumber \\
\psi&\mapsto& \psi + \mu  \nonumber \\
\mu &= &\frac{2 \nu }{\sqrt{3}} (\sqrt{2}T^{8} - T^{0}), \; \; \; \; \; \;  \nu = \sqrt{\frac{m^{2}}{4 \lambda}} .
\label{vacuum}
\end{eqnarray}

From now on, we follow the index notation established in \cite{ALVV22}. Lowercase letters from the middle of the alphabet, as i,j,k... , will designate the directions that commute and anti-commute with that of the vacuum  (\ref{vacuum}). In other words, they will represent the non-broken directions $i=(1,2,3)$ of the residual 
 $SL(2,C)$ that remains a symmetry of the vacuum after the phase transition. Accordingly, we obtained the free propagators along these directions at this phase

 \begin{eqnarray}
 <\bar{\mathcal{A}}_{\mu}^{i}\bar{\mathcal{A}}_{\nu}^{j}> &=&   (\frac{i}{k^{2}}\theta_{\mu\nu} +\frac{i\alpha}{k^{2}}\omega_{\mu\nu})\delta^{ij} ,\nonumber \\
 <\mathcal{A}_{\mu}^{i}\mathcal{A}_{\nu}^{j}> &=& - (\frac{i}{k^{2}}\theta_{\mu\nu} +\frac{i\alpha}{k^{2}}\omega_{\mu\nu})\delta^{ij} ,
 \label{prop123}
 \end{eqnarray}
 where
\begin{eqnarray}
\theta_{\mu\nu} &=& \delta_{\mu\nu}-\frac{k_{\mu}k_{\nu}}{k^{2}},\nonumber \\
\omega_{\mu\nu} &=& \frac{k_{\mu}k_{\nu}}{k^{2}} .
\end{eqnarray}
Lowercase letters from the beginning of the alphabet will designate the broken directions $a=(4,5,6,7)$, and along them we showed that
 i-particle propagators are developed 
\begin{eqnarray}
<\bar{\mathcal{A}}_{\mu}^{a}\bar{\mathcal{A}}_{\nu}^{b}> &=&   (\frac{i}{k^{2}+ig^{2}\nu^{2}}\theta_{\mu\nu} +\frac{i\alpha}{k^{2}+i\alpha g^{2}\nu^{2}}\omega_{\mu\nu}) \delta^{ab} ,\nonumber \\
<\mathcal{A}_{\mu}^{a}\mathcal{A}_{\nu}^{b}> &=& - (\frac{i}{k^{2}-ig^{2}\nu^{2}}\theta_{\mu\nu} +\frac{i\alpha}{k^{2}-i\alpha g^{2}\nu^{2}}\omega_{\mu\nu})\delta^{ab} .
\label{prop4567}
\end{eqnarray}
Along the last direction $A=8$, we found the mixed propagators   

\begin{eqnarray}
 <\bar{\mathcal{A}}^{8}_{\mu}\bar{\mathcal{A}}^{8}_{\nu}> &=& \frac{ik^{2}+ \frac{4}{3} g^{2}\nu^{2}}{k^{4}} \theta_{\mu\nu}+ \frac{i\alpha k^{2}+ \frac{4}{3}\alpha^{2} g^{2}\nu^{2}}{k^{4}} \omega_{\mu\nu} , \nonumber \\
<\bar{\mathcal{A}}^{8}_{\mu} \mathcal{A}^{8}_{\nu}> &=& \frac{4 g^{2}\nu^{2}}{3k^{4}}\{\theta_{\mu\nu} +\alpha^{2}\omega_{\mu\nu} \}, \nonumber \\
<\mathcal{A}^{8}_{\mu} \mathcal{A}^{8}_{\nu}> &=& -\frac{ik^{2}- \frac{4}{3} g^{2}\nu^{2}}{k^{4}}\theta_{\mu\nu}- \frac{i\alpha k^{2}- \frac{4}{3}\alpha^{2} g^{2}\nu^{2}}{k^{4}}\omega_{\mu\nu} . 
\label{prop8}
\end{eqnarray}

It is important to remark that these last propagators have a well studied form, playing a special role in the Wilson loop approach of fermion confinement. This property was explored in \cite{ALVV20} when we coupled this theory to fermions. At the same time, the i-particle propagators (\ref{prop4567}) promote the condensation of the three loop contribution of the object
\begin{eqnarray}
O(x) &=&{2\nu^2\over 3}\mathcal{F}^8_{\mu\nu} \bar{\mathcal{F}}^8_{\mu\nu} ,
\label{o}
\end{eqnarray}
which turns out to have a KL like propagation, characterizing a massive positive norm composite particle state.  The simultaneous  presence of both kind of fermions and gluons confinement in the same phase is another attractive achievement that this theory enables.

We found the object (\ref{o}) as part of the following invariant coycle of the symmetric phase BRST operator $s$ of (\ref{sA}), (\ref{sAbar}) and (\ref{spsi}),
\begin{eqnarray}
O(x)= Tr(\varphi \mathcal{F} \psi \bar{\mathcal{F}}) . \label{ox}
\end{eqnarray}
When the scalar fields in (\ref{ox}) attain the vev (\ref{vacuum}), $O(x)$ displays the contribution (\ref{o}) which only appears  at this broken phase. But there are contributions to (\ref{ox}) other than (\ref{o}) that do not have a well behaved propagation, and actually the dimension of the operator (\ref{ox}) extrapolates that of spacetime, even the element (\ref{o}) has one and two loop diagrams that do not lead to a KL propagation. These observations are not unexpected, as the cocycle (\ref{ox}) is characterized in the cohomology of the symmetric phase BRST operator, but we are trying to describe an observable of the broken phase. In other to evolve in this discussion we would need to discover what should be this BRST operator in the asymmetric phase. With the further developments of \cite{ALVV22}, we can now address this question.

\section{The broken phase BRST cohomology operator}

In \cite{ALVV22}, we established a recipe to access the BRST structure of the broken phase of a GUT theory. Here we will apply this scheme to the complex gauge theory of \cite{ALVV20} that we summarized in the last section. Then, following \cite{ALVV22}, our first step is to write the BRST operation (\ref{spsi}) at the transition point and define the $s_v$ operator

\begin{eqnarray}
s_v \varphi &=& ig(\varphi+ \mu) c -ig\bar{c} (\varphi + \mu)  , \nonumber \\
s_v \psi &=& ig(\psi + \mu) \bar{c}-ig c(\psi + \mu)  , \label{svpsi}
\end{eqnarray}
when the scalar fields acquire the vev (\ref{vacuum}). On the gauge fields and ghosts, $s_v$ remains unchanged as

\begin{eqnarray}
s_v \mathcal {A}_{\mu }&=&-(\partial_{\mu }c-ig[\mathcal{A}_{\mu },c]) ,\nonumber \\
s_v c &=& -igc^{2}  ,\nonumber \\
s_v \bar{\mathcal {A}}_{\mu}&=&-(\partial_{\mu }\bar{c}-ig[\bar{\mathcal{A}}_{\mu},\bar{c}]) , \nonumber \\
s_v \bar{c} &=& -ig\bar{c}^{2} . \label{svA}
\end{eqnarray}
Straightforwardly, $s_v$  is still a nilpotent operator, and becomes an exact symmetry of the action (\ref{action}) after the vacuum displacement of (\ref{vacuum}).

Now we proceed with the filtration of this $s_v$  operator in order to identify the doublet fields structure. The filter selects the linear part of the transformations (\ref{svpsi}) and (\ref{svA}). Our main interest is on the modifications brought by the vev (\ref{vacuum}) of the scalar fields at the phase transition, so we explicitly write these new linear contributions:

\begin{eqnarray}
s_0\varphi^0= \frac{a_0}{\sqrt{3}}(c^8-\bar{c}^8)   , 
\label{s0phi0}
\end{eqnarray}
\begin{eqnarray}
s_0\psi^0=- \frac{a_0}{\sqrt{3}}(c^8-\bar{c}^8)   ,
\label{s0psi0}
\end{eqnarray}
\begin{eqnarray}
s_0\varphi^8= -\frac{\sqrt{2}a_0}{\sqrt{3}}(c^8-\bar{c}^8)   ,
\label{s0phi8}
\end{eqnarray}
\begin{eqnarray}
s_0\psi^8= \frac{\sqrt{2}a_0}{\sqrt{3}}(c^8-\bar{c}^8)   ,
\label{s0psi8}
\end{eqnarray}
\begin{eqnarray}
s_0\varphi^1=s_0\varphi^2=s_0\varphi^3=s_0\psi^1=s_0\psi^2=s_0\psi^3=0,
\end{eqnarray}

\begin{eqnarray}
s_0\varphi^4= \frac{\sqrt{6}a_0}{{4}}(-ic^5-i\bar{c}^5-c^4+\bar{c}^4), 
\label{s0phi4}
\end{eqnarray}
\begin{eqnarray}
s_0\varphi^5= \frac{\sqrt{6}a_0}{{4}}(ic^4+i\bar{c}^4-c^5+\bar{c}^5) ,  
\end{eqnarray}
\begin{eqnarray}
s_0\varphi^6= \frac{\sqrt{6}a_0}{{4}}(-ic^7-i\bar{c}^7-c^6+\bar{c}^6),   
\end{eqnarray}
\begin{eqnarray}
s_0\varphi^7= \frac{\sqrt{6}a_0}{{4}}(ic^6+i\bar{c}^6-c^7+\bar{c}^7),
\end{eqnarray}
\begin{eqnarray}
s_0\psi^4= \frac{\sqrt{6}a_0}{{4}}(-ic^5-i\bar{c}^5+c^4-\bar{c}^4) ,  
\end{eqnarray}
\begin{eqnarray}
s_0\psi^5= \frac{\sqrt{6}a_0}{{4}}(ic^4+i\bar{c}^4+c^5-\bar{c}^5) ,  
\end{eqnarray}
\begin{eqnarray}
s_0\psi^6= \frac{\sqrt{6}a_0}{{4}}(-ic^7-i\bar{c}^7+c^6-\bar{c}^6) ,  
\end{eqnarray}
\begin{eqnarray}
s_0\psi^7= \frac{\sqrt{6}a_0}{{4}}(ic^6+i\bar{c}^6+c^7-\bar{c}^7)  , 
\label{s0pSi7}
\end{eqnarray}
\begin{eqnarray}
s_0c^i=s_0\bar{c}^i=s_0c^a=s_0\bar{c}^a=s_0c^8=s_0\bar{c}^8=0,
\end{eqnarray}
with
\begin{eqnarray}
a_0=\frac{2\nu i g}{\sqrt{3}}  . 
\end{eqnarray}

From (\ref{s0phi0}), (\ref{s0psi0}), (\ref{s0phi8}) and (\ref{s0psi8}) we notice that the imaginary part $c_I^8$ of the complex ghost $c^8$ is in a doublet with a linear combination of the 
scalar fields $\varphi^0$, $\psi^0$, $\varphi^8$ and $\psi^8$ . Also, the equations (\ref{s0phi4}) to (\ref{s0pSi7}) establish that the complex ghosts $c^a$ and $\bar c^a$ associated to the broken directions are in doublets with the scalars $\varphi^a$ and $\psi^a$. This is just the reflection at the BRST level of the fact that these scalar degrees of freedom migrated to the degrees associated to the i-particles appearing after the phase transition, as stated in (\ref{prop4567}).

This points us that these ghosts $c^a$ and $\bar c^a$, together with $c_I^8$, should be isolated from the BRST structure after the phase transition, as indicated in \cite{ALVV22}. In this way, we are led to the operator $s_q$

\begin{eqnarray}
s_q\mathcal{A}_{\mu }^i&=&-(\partial_{\mu }c^i+g_1f^{ijk}\mathcal{A}_{\mu }^jc^k) 
\hspace{2,1cm} s_q\bar{\mathcal{A}}_{\mu }^i=-(\partial_{\mu }\bar{c}^i+g_1f^{ijk}\bar{\mathcal{A}}_{\mu }^j\bar{c}^k) \label{sqAi}\\
s_q\mathcal{A}_{\mu }^a&=&-(g_1f^{abi}\mathcal{A}_{\mu }^bc^i+g' f^{ab8}\mathcal{A}_{\mu }^bc_R^8  ) 
\hspace{1cm}  s_q\bar{\mathcal{A}}_{\mu }^a=-(g_1f^{abi}\bar{\mathcal{A}}_{\mu }^b\bar{c}^i+g'f^{ab8}\bar{\mathcal{A}}_{\mu }^bc_R^8  ) \label{sqAa}\\
s_q \mathcal{A}^{8}_{R\mu} &=& -\partial_{\mu}c_R^8 \hspace{1cm}  s_q \mathcal{A}^{8}_{I\mu} = 0 \label{sqA8}\\
s_q\varphi^0&=& \frac{ig_1}{\sqrt{6}}\varphi^{i}(c-\bar{c})^i \hspace{1cm}
s_q\psi^0= -\frac{ig_1}{\sqrt{6}}\psi^{i}(c-\bar{c})^i\hspace{1cm}\\ 
s_q\varphi^8&=& -\frac{ig_1}{2\sqrt{3}}\varphi^{i}(c-\bar{c})^i \hspace{1cm}
s_q\psi^8= \frac{ig_1}{2\sqrt{3}}\psi^{i}(c-\bar{c})^i
\hspace{1cm} \\
s_q\varphi^i&=& \frac{ig_1}{2}\left( \frac{2\varphi_0}{\sqrt{6}}(c-\bar{c})^i+d^{8ij}\varphi^8(c-\bar{c})^j+if^{ijk}\varphi^j(c+\bar{c})^k\right)\\
  s_q\psi^i&=& \frac{ig_1}{2}\left( -\frac{2\psi_0}{\sqrt{6}}(c-\bar{c})^i-d^{8ij}\psi^8(c-\bar{c})^j+if^{ijk}\psi^j(c+\bar{c})^k\right)
    \\    s_q{ {c}}^i &=& {g_1\over{2}}f^{ijk}{c}^j{c}^k  \hspace{1cm} s_q{ \bar{c}}^i = {g_1\over{2}}f^{ijk}\bar{c}^j\bar{c}^k  \hspace{1cm} s_q c_R^8=0 . \label{sqc}
\end{eqnarray}

This is the nilpotent operator which will effectively identify the independent cohomological classes after the symmetry breaking of the complex  $SL(3,C)$ gauge theory. Among several features, we highlight the residual  $SL(2,C)$ symmetry of the new vacuum displayed by (\ref{sqAi}) with its new coupling constant $g_1$, and also the abelian symmetry with coupling $g'$ associated to the real part $\mathcal{A}^{8}_{R\mu}$ of the complex gauge component $\mathcal{A}^{8}_{\mu}$ as we see from (\ref{sqA8}). This expression also shows that the imaginary part $\mathcal{A}^{8}_{I\mu}$ now becomes a vectorial matter field. This happens because the imaginary component $c_I^8$ ceases to be a ghost of the BRST operator  $s_q$, and only the real component $c_R^8$ appears as an abelian ghost after the phase transition. This frame is actually responsible for the special mixed propagating pattern found for $\mathcal{A}^{8}_{\mu}$ in (\ref{prop8}), which in its turn is behind the development of a confining fermionic potential in the asymmetric phase of the complex theory \cite{ALVV20}. Equations (\ref{sqAa}) and (\ref{sqA8}) feature another fundamental outcome of the phase transition: they sign the braking of the holomorphicity of the complex theory. This has an amazing impact on the physics of this theory, and is ultimately responsible for the emergence of gluon condensates as composite particles. We will explain this in detail in the next section.

\section{The breaking of holomorphicity and gluon condensates} 

The set of BRST transformations (\ref{sA}) and (\ref{sAbar}), typical of a complex gauge theory, has a direct implication: the holomorphicity of the invariant gauge theory, as can be seen from the gauge sector of the action (\ref{action}). In four dimensions, there is no room for non-holomorphic pure gauge elements as long as the symmetries  (\ref{sA}) and (\ref{sAbar}) are present. The relevance of this point must be well understood. As long as non-holomorphic gauge elements are forbidden, neither there is room for i-particles condensates. Obviously, in the symmetric phase of the complex theory where holomorphicity is preserved, we do not find i-particles whatsoever. But our point is that even if the i-particles propagators 
(\ref{prop4567}) were to be found in this phase, holomorphicty would preclude i-particle condensates. Let us show why this is so. The idea of i-particle condensates is based on two main points. First, the theory must develop gauge i-particles $<\mathcal{A}_{\mu}^{a}\mathcal{A}_{\nu}^{a}>$ and conjugate i-iparticles $<\bar{\mathcal{A}}_{\mu}^{a}\bar{\mathcal{A}}_{\nu}^{a}> $ propagators, such as those in (\ref{prop4567}). Second, the theory must provide invariants with the two kind of i-particles fields $\mathcal{A}_{\mu}^{a}$ and
 $\bar{\mathcal{A}}_{\mu}^{a}$ in the same element. If both conditions are met, we see the formation of condensates as presented in figure (1).

\begin{figure}[ht]\label{fig1}
	\centering
	\begin{minipage}[b]{0.25\linewidth}
		\centering
		\includegraphics[width=0.9\textwidth]{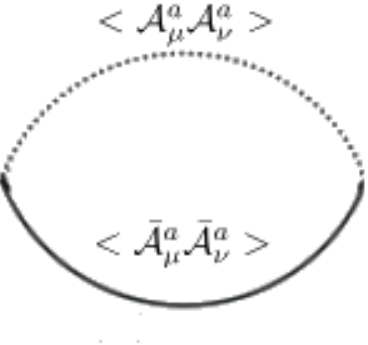}
		\label{fig1}
	\end{minipage}
\caption{Two-point condensate function.}
\end{figure}

Such structures were studied in  \cite{Bau10, Cap11, Dud10, Dud11}. From these works we know that the integral associated to the loop formed from the simultaneous presence of i-particles and conjugate i-particles propagators satisfy the KL spectral representation. This means that the condensate formed from the fields $\mathcal{A}_{\mu}^{a}$ and $\bar{\mathcal{A}}_{\mu}^{a}$ propagates as a composite particle. But in order to find  both fields in the same element, BRST must allow for non-holomorphic invariant objects, or else these composite particles will not be formed as physical observables of the theory. Now we see the necessity for the breaking of holomorphicity of the original symmetric complex field theory. And the fact is that the breaking generated by (\ref{vacuum}) leading to the BRST operator $s_q$ of (\ref{sqAa}) and (\ref{sqA8}) is also breaking holomorphicity. Then, in the broken phase, we verify the formation of several non-holomorphic non trivial elements of the $s_q$ cohomology. Some of them have the necessary ingredients to generate condensates. We list now the only three independent elements that were found
\begin{eqnarray}
\theta_0 &=& f^{8ab}\mathcal{A}_\mu^a\mathcal{A}_\nu^b f^{8cd} \bar{\mathcal{A}}_\mu^c \bar{\mathcal{A}}_\nu^d  , \nonumber \\ 
\theta_1 &=& \mathcal{A}_\mu^a\mathcal{A}_\mu^a\bar{\mathcal{A}}_\nu^b\bar{\mathcal{A}}_\nu^b , \nonumber  \\
\theta_2 &=& \mathcal{A}_\mu^a\mathcal{A}_\nu^a\bar{\mathcal{A}}_\mu^b\bar{\mathcal{A}}_\nu^b .
\label{condensates}
\end{eqnarray}

Our first observation is that all of them require a three-loop propagation structure, as in figure (2).
\begin{figure}[ht]\label{fig2}
	\centering
	\begin{minipage}[b]{0.25\linewidth}
		\centering
		\includegraphics[width=0.8\textwidth]{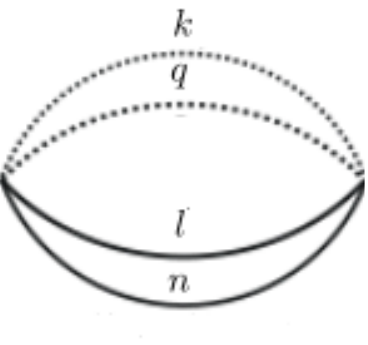}
		\label{fig2}
	\end{minipage}
\caption{Two-point condensate function with momentum $P=k+q+l+n$. Dashed lines corresponds to $<\mathcal{A}\mathcal{A}>$ and continuous lines to $<\bar{\mathcal{A}}\bar{\mathcal{A}}> $.}
\end{figure}

This turns the calculation of the KL spectral representation more involved than the one loop example just mentioned. Anyway, the particular case of the $\theta_0$ object has already been derived in \cite{ALVV20}. In fact, $\theta_0$ is the interacting element present in $O(x)$ of (\ref{o}). And, as previously argued, it is part of the invariant cocycle of the BRST operator $s$ presented in 
(\ref{ox}). This point deserves to be enlightened. This relation between the  $\theta_0$ cocycle of the broken-phase $s_q$ operator and the $O(x)$ coycle of the symmetric phase $s$ operator is not spurious. It is a product of the general theorem of cohomology which states that the cohomology of a nilpotent operator $S$ is contained inside the cohomology of any nilpotent operator obtained as a filtration of $S$ (see \cite{Sor95} for this demonstration). In our case the $s_q$ of (\ref{sqAi}) to (\ref{sqc}) can be obtained as a filtration of the $s$ in (\ref{sA}), (\ref{sAbar}) and (\ref{spsi}) on the ghosts $c^a$, $\bar c^a$, and $c_I^8$. This explains why an invariant object of the asymmetric phase, $\theta_0$ , is related to another invariant $O(x)$ of the symmetric phase. This has a physical implication as we understand that observables of a phase of a theory can be found as part of observables of a previous phase. In \cite{ALVV20} we found a condensate with the right KL propagation only by inspection, and the connection with an observable of the broken phase of the theory was just guessed by identifying it as part of an observable of the symmetric phase. Here we see why this was possible, and prove how the condensate $\theta_0$ is in fact an observable. 

The other two candidates, $\theta_1$ and $\theta_2$, are only being identified now after studying the cohomology of $s_q$. It is not difficult to see that they will also contribute to the condensate once we write the integral of the three-loop graph,where we consider $(\alpha=1)$ in (\ref{GF}), describing their propagation

\begin{eqnarray}
	I(\theta_1) & = &\frac{2}{\pi}\int \frac{d^{4}k}{4\pi^2} \frac{d^{4}q}{4\pi^2} \frac{d^{4}n}{4\pi^2} 
	\frac{d^{4}l}{4\pi^2}
	\delta(n-[p-q-k-l])
	<A^{a}_{\mu}(k)A^{d}_{\rho}(-k)>
	<A^{a}_{\mu}(q)A^{d}_{\rho}(-q)> \nonumber \\ 	
	&&	<\bar{A}^{c}_{\nu}(l)\bar{A}^{b}_{\sigma}(-l)>
	<\bar{A}^{c}_{\nu}(n)\bar{A}^{b}_{\sigma}(-n)> \nonumber \\ 
	&=& 
	\frac{2 }{(\pi)^9 }\int  {d^{4}k}{d^{4}q}{d^{4}n}{d^{4}l}\delta(n-[p-q-k-l]) \nonumber \\ &&
	\frac{1}{k^2-i{m^2}} \frac{1}{q^2-i{m^2}}  \frac{1}{l^2+i{m^2}} \frac{1}{n^2+i{m^2}},\nonumber \\
           I(\theta_2) & = & \frac{1}{4} I(\theta_1) .
	\label{srigbill}
\end{eqnarray}
A general proof stated in \cite{Bau10} establishes that this integral leads to a KL propagation, allowing to interpret $\theta_1$ and $\theta_2$ as condensates. Even more, we can see that this integral displays the same integrand of that calculated in \cite{ALVV20} for $\theta_0$. In this way, we understand that all three objects in (\ref{condensates}) contribute to the same particle state.

Another point that we must call attention is that $\theta_1$ and $\theta_2$ also lead to non-trivial elements in the cohomology of the BRST operator $s$ of the symmetric phase. They are respectively

\begin{eqnarray}
O_1 &=&  [ Tr(\mathcal{D}_{\mu}\varphi)(\mathcal{D}_{\mu}\psi)]^2  \, ,  \label{o1} \\
O_2 &=&  [ Tr(\mathcal{D}_{\mu}\varphi)(\mathcal{D}_{\nu}\psi)] [ Tr(\mathcal{D}_{\mu}\varphi)(\mathcal{D}_{\nu}\psi)] \, . 
\label{o2}
\end{eqnarray}

It is interesting to notice that $O(x)$, $O_1$ and $O_2$, although belonging to the symmetric phase cohomology, cannot be present in the symmetric phase action (\ref{action}) as they extrapolate the four dimensional bound. No term in  (\ref{action}) is actually the origin of any condensate in the broken phase in this theory.

\section{Conclusion}

In this work we applied the recent development of the concept of a BRST operator of the asymmetric phase of a spontaneous broken field theory \cite{ALVV22} in the cohomological characterization of condensate observables. Here we studied the complex field theory introduced in \cite{ALVV20}. We showed how the condensate $\theta_0$ in (\ref{condensates}), presented for the first time in \cite{ALVV20} is associated to a non-trivial cohomological class of the nilpotent BRST operator $s_q$ ((\ref{sqAi}) to (\ref{sqc})). Therefore, we establish it as a BRST observable of the broken phase of the complex field theory. Furthermore, we found out here two other independent non-trivial classes associated to $\theta_1$ and $\theta_2$ . As these three objects in (\ref{condensates}) have a KL propagation given by the same integral (\ref{srigbill}), we find that they all contribute to the same condensed particle state.

Another observation is that the origin of these elements is traced to non-trivial cocycles ( (\ref{ox}), (\ref{o1}) and (\ref{o2}) ) of the BRST operator of the symmetric phase written in (\ref{sA}), (\ref{sAbar}) and (\ref{spsi}). The  interesting point here is that they are all objects which extrapolate the dimensionality of the four dimensional space-time. As an open question is still the immersion of the starting action (\ref{action}) inside a larger theory where the symmetry breaking mechanism could evolve, it is tempting to imagine that a higher dimensional theory could be behind this scenario.

\end{document}